\documentclass{PoS}
\usepackage{listings}
\usepackage[T1]{fontenc}
\usepackage[utf8]{inputenc}
\usepackage{dirtree}
\usepackage{caption}
\usepackage{subfig}

\title{A Prototype Data Format for the Cherenkov Telescope Array: Regions Of Interest (ROI)}

\ShortTitle{ROI}

\author{
    \speaker{Ramin Marx}\\
    Max-Planck-Institut für Kernphysik, PO Box 103980, 69029 Heidelberg, Germany\\
    E-mail: \email{ramin.marx@mpi-hd.mpg.de}
}

\author{
    Raquel de los Reyes\\
    Max-Planck-Institut für Kernphysik, PO Box 103980, 69029 Heidelberg, Germany\\
    E-mail: \email{raquel.de.los.reyes@mpi-hd.mpg.de}
}

\abstract{
    The Cherenkov Telescope Array (CTA) is a ground-based $\gamma$-ray observatory that will observe the full sky
    in the energy range from 20 GeV to 100 TeV from facilities in both hemispheres.
    It is proposed to consist of more than 100 telescopes and the large amount of data produced will exceed the volume of current VHE Imaging Atmospheric Cherenkov Telescopes by $\sim$two orders of magnitude.
    This volume of data represents a new challenge to the community, which is looking for new data formats to transfer and store the CTA data.
    One of the prototypes currently under study is the ROI (Regions Of Interest) file format for camera images.
    It can store only those pixels of a camera image that are close to the shower, thus removing the major part of the night sky background (NSB) while keeping all pixels that might belong to the shower.
    Simple on-the-fly compression is used to reduce the file size even further.
    Here, we explain the ROI prototype in detail and present preliminary results when applied to simulations.
}

\FullConference{
    The 34th International Cosmic Ray Conference,\\
    30 July- 6 August, 2015\\
    The Hague, The Netherlands
}

\begin{document}

\section{Introduction}
CTA will be the first open observatory of very-high-energy $\gamma$-rays.
It will be the successor of the current generation of ground-based imaging atmospheric Cherenkov telescope (IACT) experiments.
Arrays constisting of up to 100 telescopes may combine up to six types of telescope and up to seven types of cameras.
This design exceeds the dimension and complexity of the current IACT experiments, which are formed by a maximum of 5 telescopes (H.E.S.S.~\cite{HESS}) and with no more than two telescope and camera types.
The telescopes will record Cherenkov light coming from the extensive air showers (EAS) produced by primary $\gamma$-rays and, mostly, by cosmic rays (CR).
The high expected trigger rates of several tens of kHz, together with the $\sim 10^3$ to $10^4$ pixels per camera, will lead to huge amounts of data in form of camera images of air shower events~\cite{DATA}.

The ROI (Region Of Interest) file format has been developed in order to store these images in an efficient and simple way.
It can store full camera images and pixel lists, like other formats, but also the ROI of the camera image, which is the part that contains the shower.
This allows for a compromise between storing all data and only the significant pixels.
Storing all information of all pixels would result in too much data, whereas storing only the pixels that are assumed to belong to the shower might throw away important information needed for gamma/hadron separation, for example.
Instead, a ROI around the shower is stored that includes all significant pixels as well as the lower intensity shower pixels and the NSB pixels.
Since there are cases where full camera images or pixel lists are needed, these are also supported by the ROI format.
Furthermore, it reduces the resolution of the intensities and times of maximum by ignoring non-significant digits, which saves space and allows to store these values as integers.

At the moment, ROI has been applied and tested only for storing calibrated camera images, but the format could be extended to support raw data as well.

\section{The Region Of Interest}
\begin{figure}
    \includegraphics[width=\textwidth]{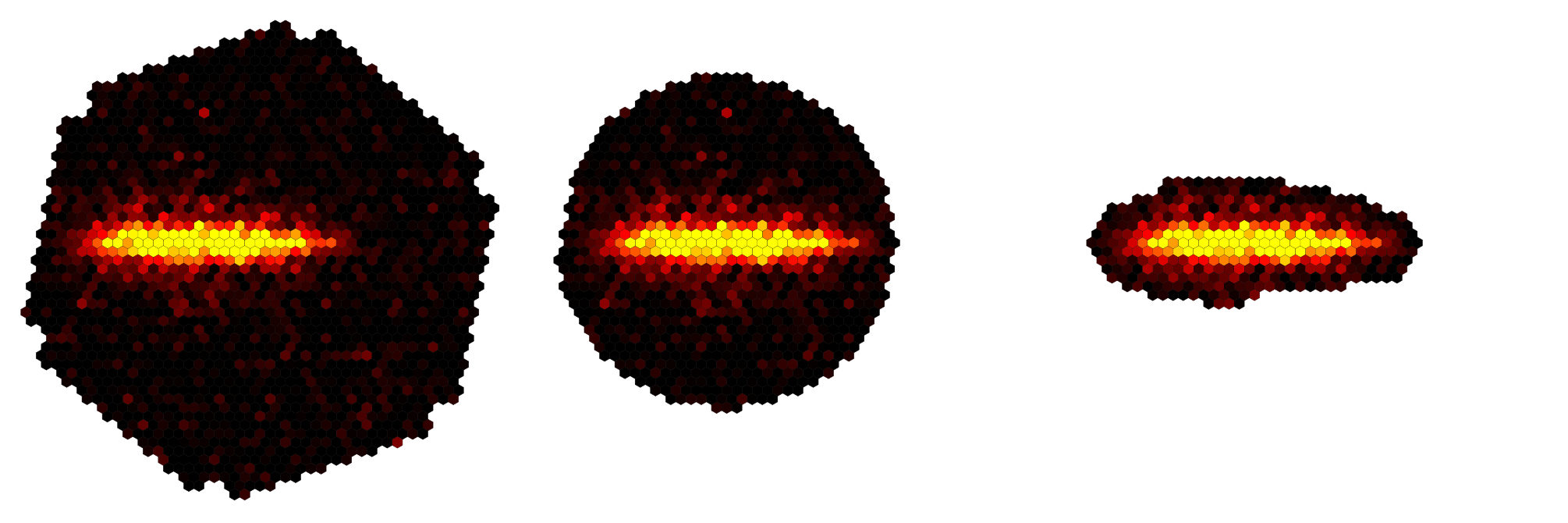}
    \caption{Full camera image (left), ROI (centre) and pixel list (right).}
    \label{fig:camroilist}
\end{figure}

The ROI of a shower image is defined as the smallest circle around the centre of gravity of the cleaned image that contains all pixels that survived the image cleaning (see figure~\ref{fig:camroilist}).
This implies that the ROI depends strongly on the image cleaning, so the NSB level and the electronic noise of the camera must be known well.
In order to lose as few significant pixels as possible, some extra rows of neighbours should be included in the cleaned image.
Since the ROI would include them as well, it is clear that the ROI never loses any pixels that are included in the cleaned image.
On the contrary, the ROI even includes pixels below the cleaning threshold and outside of the neighbourhood of the pixel list, so the probability that all shower and subshower pixels are captured is higher than if the pixel list was used.
Figure \ref{fig:camroilist} shows that the ROI includes some pixels that look like NSB pixels, but might be shower pixels.
However, the pixel list does not include these pixels, although the list includes three rows of neighbour pixels.
These additional pixels can be used for improved reconstruction~\cite{impact} or for calibration.
They can also be useful if a later calibration results in different cleaning thresholds.
The circular shape was chosen because it is easy to parameterise and because it matches the form of the lower energy showers, which dominate due to the typical proton spectrum of ~-2.7.

\section{Structure}
A ROI file has no header, it only contains a concatenation of (calibrated) camera events.
Each event data block starts with the event header, containing the event tag, the type of the event, global event id, global timestamp and the number of participating telescopes.
Finally, the data blocks with the telescope events follow.
Each telescope event starts with the telescope event header, containing the telescope id, the storage type (full camera, ROI or list), the lowest pixel intensity in the shower $I_{min}$ and
the intensity threshold for time-of-maximum storage $I_T$ (for all pixels with $I > I_T$, the time of maximum of the Cherenkov pulse is stored - this way, no ids are needed for the pixels whose time of maximum is stored).
Then, the pixel intensities and the times of maximum follow.
The storage type defines what procedure is applied to store the pixel intensities and their times of maximum:
\begin{itemize}
    \item If the full camera has to be saved, there is no need to store the pixel indices, so only the pixel intensities and times of maximum are stored.
        This storage type is appropriate for calibration events, large showers that cover most camera pixels and very small showers that vanish, if image cleaning is applied.
    \item The ROI does not require pixel indices either, since the circle defines which pixels are included.
        This storage type is appropriate for most showers.
    \item List storage requires the pixel indices and can be used for storing muons, for example.
\end{itemize}
The intensities and the times of maximum of the pixels selected by one of the storage methods above are then compressed and written to file.
However, before a pixel intensity $I$ is stored, it must be checked if $I$ is defined, because procedures like image cleaning can remove pixels and replace them with a value signaling that the intensity is not defined.
In such a case, a zero byte is written and the next pixel is processed.
Otherwise,
\begin{eqnarray*}
    J = \textrm{round}\left(\frac{I - I_{min}}{pe\_res}\right)
\end{eqnarray*}
is stored.
By substracting $I_{min}$, the expression is guaranteed to be positive.
Then dividing by $pe\_res$ (the photoelectron resolution of the camera) and later rounding, an integer $J$ is obtained that holds the significant information of $I$, while the insignificant information is removed.
In order to map the minimum intensity in the shower to a different value than an undefined intensity (which is already mapped to 0), $I_{min}$ is decremented by 1,
    before the encoding of the telescope event begins, ensuring that all $J > 0$.
Depending on its value, $J$ is stored with a different number of bytes:
If $J < 255$, one byte is enough to store $J$.
If $255 \leq J < 65535$, a $255$ is written to signal that one byte is not enough, following two bytes containing the value of $J$.
If $J \geq 65535$, a $255$ is written to signal that one byte is not enough, following a $65535$ to signal that two bytes are not enough, following four bytes containing the value of $J$.

If $I > I_T$, the time of maximum $T$ for that pixel is stored as well.
The same encoding scheme as for the pixel intensity is used:
\begin{eqnarray*}
    J = \textrm{round}\left(\frac{T - T_{min}}{tom\_res}\right),
\end{eqnarray*}
with $T_{min}$ being the lowest time of maximum in the shower and $tom\_res$ the time-of-maximum resolution of the camera.

Since integers do not contain any noise in the unused bits, storing them instead of floating point numbers reduces the entropy in the resulting file,
    allowing entropy compressors like \textsc{ZIP} to reduce the size even further.

\section{Data reduction}

First, the two methods for data reduction in ROI (selecting a subset of pixels from the the full camera image and reducing the resolution) are quantified.

\subsection{Storage type}

\begin{figure}[ht!]
    \centering
    \subfloat[][Hillas width]{\includegraphics[width=\textwidth]{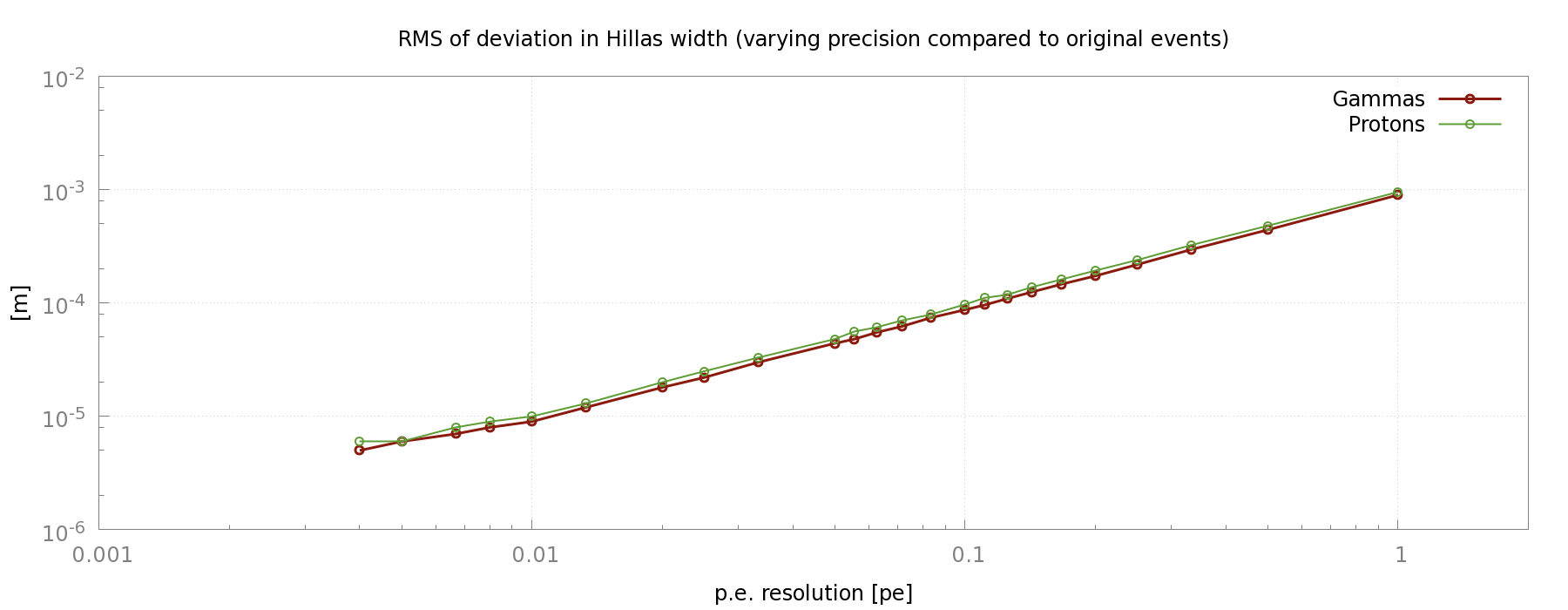}}
    
    \subfloat[][Hillas orientation]{\includegraphics[width=\textwidth]{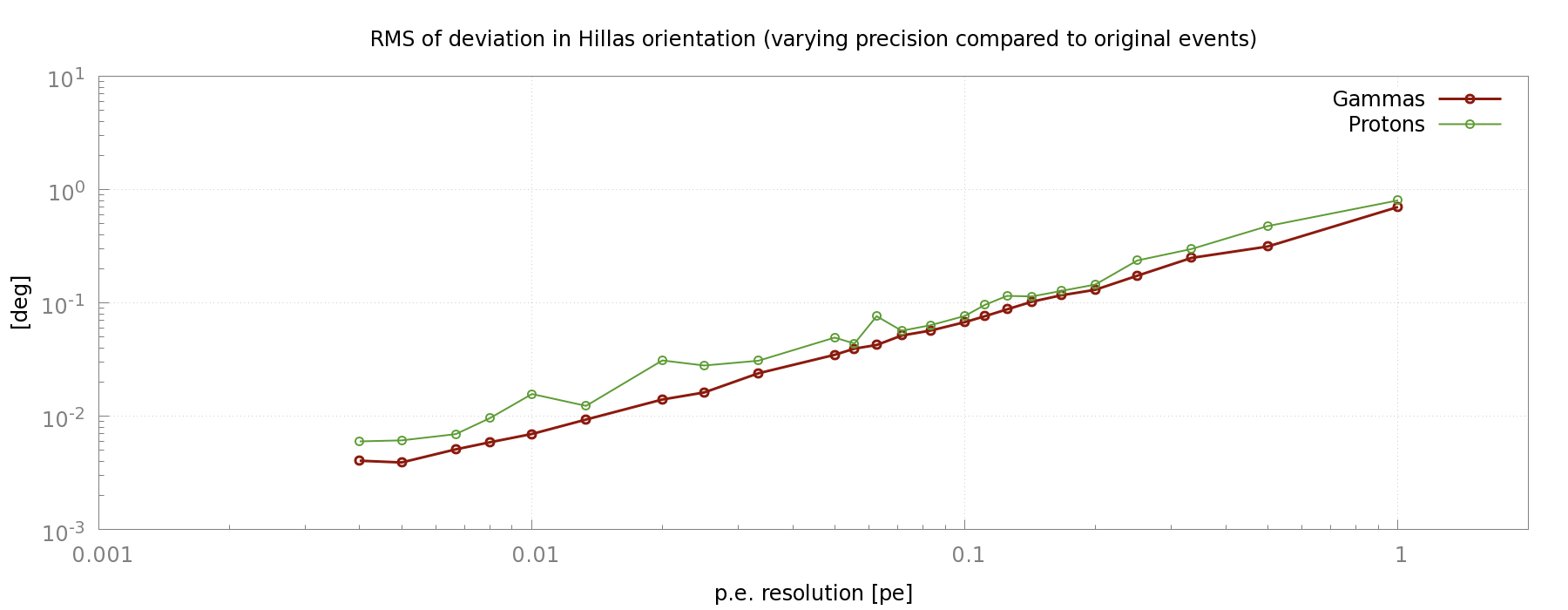}}

    \subfloat[][Event size]{\includegraphics[width=\textwidth]{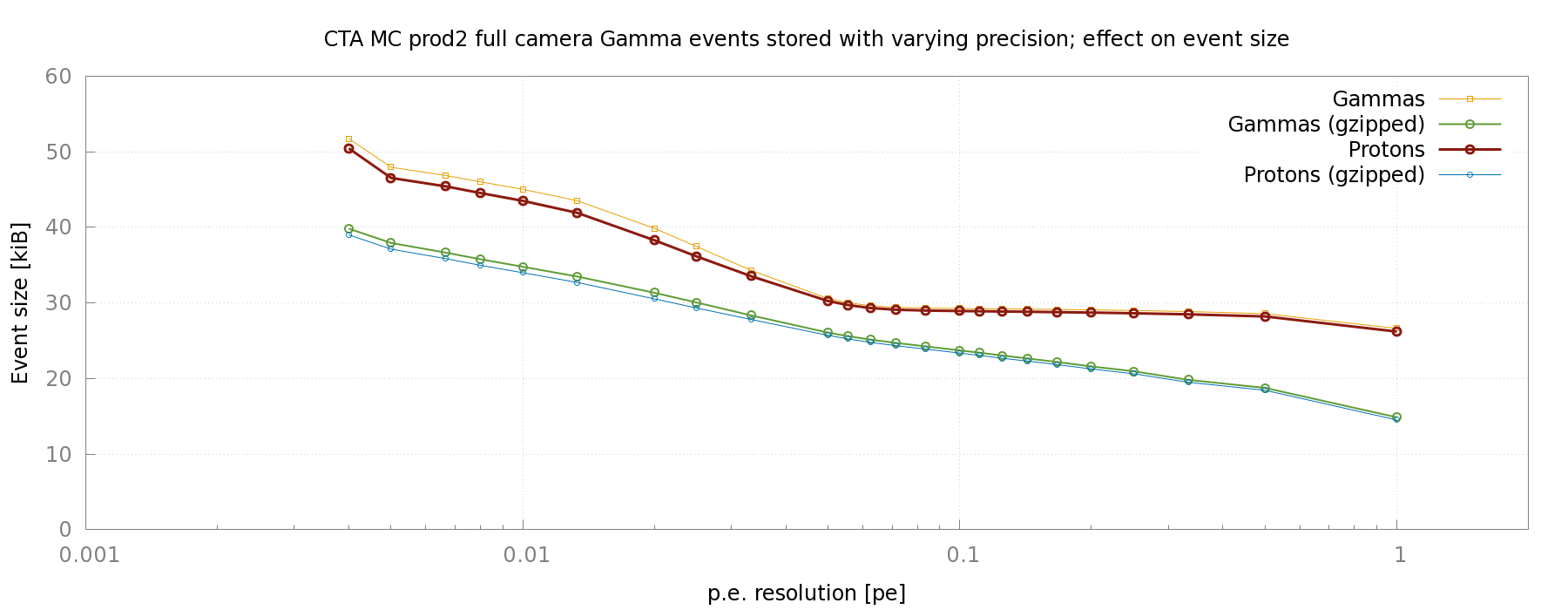}}
    
    \caption{(a) Deviation in Hillas width gammas, (b) deviation in Hillas orientation and (c) average event sizes.}
    \label{fig:plots}
\end{figure}

For this test, 1888 CTA MC proton test showers~\cite{MC}, created by \textit{simtelarray}~\cite{simtelarray} were written to a ROI file.
Protons are a realistic example, because the majority of showers in the real CTA data will be protons.
Stored are either full camera images, pixel lists of cleaned images or ROI around these pixel lists.
For this exercise, the photoelectron resolution and the time-of-maximum resolution of the camera are fixed to $pe\_res = 0.0625$ pe and $tom\_res = 0.1$ ns
    and the time of maximum $T$ is saved for all pixels with an intensity above $I_T = 0$ pe.
According to figure \ref{fig:plots} (a) and (b), these values are reasonable.

The image cleaning keeps only those pixels that have an intensity $> 10$ p.e. (photo electrons) and at least 3 neighbours with an intensity $> 5$ p.e.
After cleaning, three rows of neighbours around the remaining pixels are added.
If the image cleaning removes all pixels, the whole camera image is stored, as a precaution against losing any data.

Due to the different telescope types in CTA, the image cleaning will in reality have to be optimised for the NSB level and the pixel size for each camera.
For simplicity, the same image cleaning algorithm and values were assumed for the full CTA array.
In case of the image cleaning accidentally removing all pixels, the whole camera is saved, which guarantees that no data will be lost.


%

\begin{table}[ht!]
    \centering
\begin{tabular}{|l|r|r|r|r|r|}
    \hline
    Storage type & $F_{full\_cam}$ & $N_I$ & $N_T$ & event size & $S_{full\_cam}$ \\
    \hline
    Full camera & 100\% & 19421 & 7710 & 30.0 kiB & 100\% \\
    ROI         &   9\% &  3951 & 1945 &  6.9 kiB &  29\% \\
    List        &   9\% &  1917 & 1007 &  4.3 kiB &  46\% \\
    \hline
\end{tabular}
\begin{tabular}{|l|r|r|r|r|r|}
    \hline
    Storage type & $F_{full\_cam}$ & $N_I$ & $N_T$ & event size & $S_{full\_cam}$ \\
    \hline
    Full camera & 100\% & 19421 & 7710 & 30.0 kiB & 100\% \\
    ROI         &   0\% &  2994 & 1567 &  5.4 kiB &   0\% \\
    List        &   0\% &   760 &  537 &  2.6 kiB &   0\% \\
    \hline
\end{tabular}
\caption{Comparison of the three different storage methods (full camera, ROI and pixel list):
    $F_{full\_cam}$ is the fraction of full camera images among all telescope events in the file,
    $N_I$ is the number of intensities stored per event (number of pixels whose intensity is stored; sum across all telescopes),
    $N_T$ the number of times-of-maximum stored per event (number of pixels whose time of maximum is stored; sum across all telescopes),
    average event size and
    $S_{full\_cam}$ the impact of the full camera images on the file size (the fraction of the file size that is consumed by full camera telescope events).
    Top: ROI and List storage with 9\% full camera events - this is a bad case, in which image cleaning erases all pixels in one telescope event out of ten.
    Bottom: ROI and List storage without any full camera events - this is the ideal case, in which image cleaning always returns some pixels.}
\label{tab:results_selection}
\end{table}

As can be seen in the upper table \ref{tab:results_selection}, ROI and pixel lists consume significantly less space than full camera images (6.9 kiB and 4.3 kiB compared to 30.0 kiB),
    because the numbers of pixel values stored per event ($N_I$ and $N_T$) are much lower (3951+1954 and 1917+1007 compared to 19421+7710).
Due to non-optimal calibration and image cleaning (thresholds), 9\% of the telescope events vanish during image cleaning (needed for \textit{ROI} and \textit{List} storage)
    and are therefore stored as full camera images, in order not to lose any data.
The now included full camera events increase the number of pixel values stored per event significantly, which is also reflected in $S_{full\_cam}$:
    only 9\% of the telescope events are stored as full camera images, but they are responsible for 29\% / 46\% of the file size (ROI/pixel list).
The lower table shows the results for an idealised scenario with improved calibration and image cleaning, in which no full camera telescope events need to be stored (and therefore $F_{full\_cam} = S_{full\_cam} = 0$).
Only (2994 + 1567)/(19421 + 7710) = 16\% of the pixel values need to be saved for \textit{ROI} storage, resulting in an average event size of 5.4 kiB/30 kiB = 18\% of the average event size of the full camera storage.
For \textit{pixel list} storage, (760 + 537)/(19421 + 7710) = 4.7\% of the pixel values need to be saved, resulting in an average event size of 2.6 kiB/30 kiB = 8.6\% of the average event size of the full camera storage.
Note that since for a ROI, the pixel indices do not need to be stored, an event has only twice the size of an event stored as list, but it contains 3.5 as many pixel values.

\newpage

\subsection{Testing the Effect of Reduced Resolution}
The idea of this test is to check if a reduced resolution has an impact on the image parameters Hillas width and Hillas orientation, which are important for gamma/hadron separation and direction reconstruction.
The test files are CTA MC prod 2 files in \textit{simhess} format: a proton file containing 1888 events and the gamma file containing 5585 events.

Both files are converted to ROI files using full camera storage and with a varying p.e. resolution of the camera (from 0.004 to 1 p.e.), which means that the pixel intensities are stored with varying accuracy.
For example, a p.e. resolution of 0.1 means that the digits after the first comma are not signficant and do not need to be stored.
The resolution of the time of maximum is fixed to 0.1 ns.

The test procedure is the following:
First, store the events with highest p.e. resolution to file $A$.
Then, for each p.e. resolution in the test interval (0.004 - 1 p.e.), apply it for all events and store them to file $B$, read the original file $A$ and compute the RMS of its deviation from $B$ in terms of Hillas parameters.

Figures \ref{fig:plots} (a) and (b) show these plots for Hillas width and Hillas orientation.
As the RMS of Hillas orientation itself is typically $\sim$ 10 degrees, the errors introduced by the reduced resolution are very small and can be neglected.
The same applies for the error in Hillas width, because the RMS of the Hillas width is $\sim$ 12 cm.
The average event size decreases significantly with the p.e. resolution (see \ref{fig:plots} (c)), until it stagnates at 0.06 p.e.
There, the quantisation let the majority of the pixel intensities end up in the dynamic range of one byte, which is the smallest possible range in ROI.
The ZIP algorithm, however, is capable of exploiting the increasing number of zeros and reduces even those event sizes in the saturated regime.
Note that the plots show no noteworthy differences between gammas and protons.

\section{Conclusion}
The ROI format is proposed as a data format for calibrated CTA events.
The algorithm guarantees that all pixels that are included in the cleaned image are preserved in the ROI, so no significant pixels are lost.
However, for storing ROI or pixel lists, accurate image cleaning is mandatory, like for all other methods that store only a selection of pixels.

It has been shown that the inherent data reduction is considerable and that it has no measurable impact on the reconstruction quality.
Currently, tests are done to check if the extra pixels in the ROI can be used for improved gamma/hadron separation.
These promising results draw interest to investigate if it is feasible to store raw data in the ROI format as well, because full camera images require too much space and pixel lists might not include pixels that should be included after better calibration.
For the final results on this data format, evaluation tests within the full analysis chain will be performed within CTA, to verify that there is no performance degradation.

Since the algorithms are very simple, the IO routines are just 250 lines of code, without having any external dependencies.
This makes it easy to understand and easy to maintain.
The code is part of the MESS software framework~\cite{MESS} and can be downloaded from

http://www.mpi-hd.mpg.de/\textasciitilde rmarx/mess.

\end{document}